\documentclass[12pt]{iopart}
\usepackage{iopams,amssymb,hyperref,graphicx}

\begin{document}
\title{Power spectra of TASEPs with a localized slow site}

\author{L. Jonathan Cook$^{1,2}$ and J. J. Dong$^3$}

\begin{abstract}
The totally asymmetric simple exclusion process (TASEP) with a localized defect is revisited in this article with attention paid to the power spectra of the particle occupancy $N(t)$. Intrigued by the oscillatory behaviors in the power spectra of an ordinary TASEP in high/low density phase(HD/LD) observed by Adams \emph{et al.} (2007 \textit{Phys. Rev. Lett.} \textbf{99} 020601), we introduce a single slow site with hopping rate $q<1$ to the system. As the power spectrum contains time-correlation information of the particle occupancy of the system, we are particularly interested in how the defect affects fluctuation in particle number of the left and right subsystems as well as that of the entire system. Exploiting Monte Carlo simulations, we observe the disappearance of oscillations when the defect is located at the center of the system. When the defect is off center, oscillations are restored. To explore the origin of such phenomenon, we use a linearized Langevin equation to calculate the power spectrum for the sublattices and the whole lattice. We provide insights into the interactions between the sublattices coupled through the defect site for both simulation and analytical results.
\end{abstract}

\address{1.  Department of Physics, Virginia Tech, Blacksburg, VA 24061, USA\\
2.  Department of Physics and Engineering, Washington \& Lee University, Lexington, VA 24550, USA\\
3.  Department of Physics, Hamline University, Saint Paul, MN 55104, USA}

\ead{lacook1@vt.edu, jdong01@hamline.edu}



\section{Introduction}
\ \ \ \ There has been increasing interest in the investigation of complex phenomena, most prominently in problems inspired by biology in which a network of reactions are coordinated with surprising efficiency and accuracy. At the heart of these problems, however, are smaller fundamental ingredients and various pathways. Physics is notably well-suited for approaching such complex problems from a "reductionist" perspective: Using simplified models to capture the essential constituents and describe the fundamental behaviors. One of the paradigmatic systems in non-equilibrium statistical mechanics is the totally asymmetric simple exclusion process(TASEP). Using a matrix-product formalism\cite{DEHP, EvansRev} or recursion relations\cite{Schutz}, the steady state of TASEP with open boundaries can be obtained exactly. TASEP has been applied to model both transcription (\cite{KlumppTSX,Tripathi}) and translation(\cite{MG,LBSZia,TomChou,Thiel09}) processes in protein synthesis as well as molecular motors(\cite{Motor,motor_rev}), vehicular traffic(\cite{Chowdhury,Popkov}) and inhomogeneous growth processes (\cite{KPZ,WolfTang}). Computationally affordable, TASEP also shines light on properties of complex systems that are experimentally unattainable at the moment. The steady state properties of TASEP have been extensively studied both in the abstract and with applications. Despite its simple construction, TASEP displays non-trivial behaviors from a theoretic point of view such as phase transitions in one-dimension with short-range interaction.  Recent investigations on TASEP in frequency space \cite{Dudzinski,NAS, Takesue,PPOF,deGierPRL,deGier,AZS,Gupta,CZ} provide insights into the rich dynamic behaviors of this model. Of more relevance to this article, the authors of \cite{AZS,CZ} look into the power spectrum of the time-dependent particle occupancy of TASEP and uncovered even more surprises when oscillations are observed for TASEP in high/low density phases.

In this article, we focus on the power spectrum of TASEP with \emph{local inhomogeneity}, especially the interaction between the two sublattices separated by the inhomogeneity. We outline the biological motivation in Section \ref{sec:motivation} and review previous works pertaining to this study in Section \ref{sec:rev}. We define our model and provide simulation results in Section \ref{sec:sim}. Using Langevin dynamics, we present our theoretical results which provide insights into the interactions between the sublattices in Section \ref{sec:theo}. We conclude in Section \ref{sec:sum}

\subsection{\label{sec:motivation}Biological Motivation}
\ \ \ \ In a living organism\cite{cell}, proteins are not only the building blocks of cells (comprising most of a cell's dry mass), 
but they perform nearly all cellular functions as well. To optimally synthesize proteins, namely first  \textit{transcribing}  the genetic information from DNA into a messenger RNA (mRNA) and later \textit{translating} mRNA into a functional protein, is of crucial value \textit{in vivo} for the organism and  \textit{in vitro} for pharmaceuticals. We devote our attention to the \textit{translation} process here.

The template of translation, mRNA, is composed of nucleotide triplets called ``codons'' that contain information for the desired amino acids. As there are four distinct nucleotides(A, U, C, G), there are 64 codons. Except for three stop codons(UAA, UGA, and UAG)  that signal the termination of translation, the rest ``code'' for one of the twenty amino acids with degeneracy ranging from 1 (e.g. AUG for methionine and UGG for tryptophan) to 6 (e.g. CGN and AGR for Arginine).
In prokaryotes such as a bacterium,  translation initiates when ribosomes -- large molecules acting as an assembling machinery -- encounter the start codon AUG. During the elongation stage, ribosomes move along the mRNA and incorporate transfer RNAs(tRNA) with correct amino acids attached. Signaled by one of the three stop codons, translation terminates and the amino acids are later joined by peptide bonds to form different sequences and, with proper folding, result in various final protein products.

Codons coding for the same amino acid are termed ``synonymous.'' Even among synonymous codons, different tRNAs may be employed. Therefore all codons are translated, in principle, at a different rate depending on the availability of the cognate tRNA's. When a certain tRNA becomes scarce or even unavailable, the ribosome awaits on the corresponding codon (``slow codon''). This slows down the translation process and can lead to translation error or frame-shifting which potentially results in an erroneous protein product or terminates translation prematurely.
In light of the tRNA cellular concentrations in \textit{Escherichia coli}\cite{DNK} which can differ by nearly an order of magnitude even among synonymous codons, we look into the effect of having \emph{one} slow codon at different locations along the mRNA.  Exploring the power spectrum of such modied TASEPs is expected to provide information on the fluctuation in the timescale of particle transport through a finite system, a feature of biological relevance. 

\subsection{\label{sec:rev}Review of previous works}
\ \ \ \  TASEP comprises of a one-dimensional (1D) lattice of $\mathbb{L}$ sites and particles of size 1 in units of lattice sites moving uni-directionally with hard-core exclusion. The steady state properties of an ordinary TASEP with open boundaries, with entry/exit rates $\alpha/\beta$ and particle hopping rate $\gamma=1$, have been extensively studied(\cite{DEHP,EvansRev,Schutz}). Given different $\alpha$ and $\beta$, the system can be found in a low density ($\alpha<\beta, \alpha<1/2$, ``LD''), high density ($\beta<\alpha, \beta<1/2$,``HD''), maximal current ($\alpha>1/2, \beta>1/2$, ``MC'') and a shock phase ($\alpha=\beta<1/2$,``SP''). 

It is not until recently that the dynamics of TASEP are getting more attention\cite{Dudzinski,NAS, Takesue,PPOF,deGierPRL,deGier,AZS,Gupta,CZ}. In \cite{PPOF}, a combination of 
domain wall theory and Boltzmann-Langevin theory well characterized the dynamics of the LD and HD phases. Soon after, the complete power spectrum of the transition matrix of a partially asymmetric exclusion process was obtained using  the Bethe ansatz\cite{deGierPRL,deGier}, providing an accurate description of the short-range particle-hole correlations. 
The authors of \cite{AZS} investigated the effects of system sizes 
contained in the power spectra of the time series of system occupancy $N(t)$, which provides insight into the time-correlation of $N(t)$ and how long it takes the noise to propagate through out the entire system. Defining a Fourier transform $\tilde N(\omega)\equiv \sum_{t=1}^{T} N(t) \e^{i\omega t}$ where $\omega = 2\pi m/T$, $T$ is the length of each run and $m \in [0,T]$, the authors obtained the power spectrum by averaging over 100 such realizations:
\begin{equation}
I(\omega)=\langle |\tilde N(\omega)|^2 \rangle.
\label{eq:pow}
\end{equation}

When looking at the HD and LD phases, they found marked oscillations 
of $I(\omega)$ that damp into power laws. They further explored the origin of such behavior by taking the 
continuum limit and using a stochastic equation of motion. Their analyses and simulation results
reveal the oscillatory minima capture the system size and are located at $n\cdot 2\pi v/\mathbb{L} $ (or $n \cdot vT/\mathbb{L}$ in $m$) where $n$ is an integer, $v = 1-2\overline{\rho}$ and $\overline{\rho } $ is the average bulk density.  Note, the quantity $v/\mathbb{L}$ is simply the time the fluctuation takes to traverse the lattice.

Their results are useful in that many physical systems, e.g. mRNA with several hundred codons, are far from the thermodynamic 
limit. The timescale at which noises traverse the entire system can be significant for particle transport. In the case of cellular protein synthesis where multiple mRNA's are translated simultaneously, knowing the finite size effect on the fluctuations of total occupancy could shine some light on how mRNA competes for the ``translation machinery,'' i.e. the ribosomes. 

Our study extends the investigations in \cite{AZS} to studying the power spectra of TASEP with one localized defect (or ``slow codon'' in biological sense) where the hopping rate $q<\gamma=1$. Since an mRNA consists of a series of \emph{completely} inhomogeneous $\gamma$'s, studying the effect of \emph{one} defect to the overall particle throughput serves as an ideal starting point.
We explore both the entire system and the two subsystems joined by the defect. Our simulations show that when the defect is off the center of the lattice, the power spectrum of the entire system reflects such ``bias'' while each sublattice can still be characterized by modifying the theory for an ordinary TASEP. When located in the middle of the lattice, the cross correlations between the sublattices eliminate the oscillations, leaving a power law decay.  The continuum theory in \cite{AZS} cannot be readily modified to account for this case, so we opt to use discrete space and time units to analytically explore the power spectra for all locations of the slow site.
Our results furthered the work by Adams \emph{et al.} \cite{AZS} in the following sense: a)Since the defect can be located anywhere throughout the lattice, the left and right subsystems can have different sizes and therefore a different timescale for a hole-particle pair to propagate through. We provide both simulation and analytical results to investigate such difference; b) The left and right subsystems are coupled through the defect, leaving this site and its rightmost neighbor crucial in theoretical treatment. More careful considerations are needed which we present in the later section of this article.

\section{\label{sec:sim}Model and simulation results}
\ \ \ \ We define our system on a 1D lattice of $\mathbb{L}$ sites with open boundaries. $\mathbb{L}=1000$ for most simulation data unless otherwise specified. Particles, injected from the left end with rate $\alpha$ and taken out at the right end with $\beta$, move along the lattice with $\gamma=1$ \emph{except} at the $k^{\textrm{\tiny{th}}}$ site where a defect is introduced with hopping rate $q<\gamma$. Using a mean-field approximation\cite{DSZ06}, one finds when the defect site is far from the system boundaries, the left and right sublattices can be viewed as two TASEP's joined respectively by exit and entry rate of
\begin{equation}
q_{\textrm{\tiny{eff}}}=\displaystyle\frac{q}{1+q} \approx q \textrm{   when $q \ll 1$}.
\label{q_eff}
\end{equation}
To ensure a flat density profile without loss of generality,  $\alpha$ and $\beta$ are chosen to equal $(1- q_{\textrm{\tiny{eff}}})$.

We use a random sequential updating scheme and define one Monte Carlo step (MCS) as making $\mathbb{L}+1$ attempts to update the lattice, giving on average one chance to update each lattice site as well as introduce a new particle. All measurements are taken after 10K MCS to ensure the system has reached steady state. $N(t)$ is collected every 10 MCS for $t=1,2, ... T$ where $T=10^4$($10^5$ MCS) for most of the data. We adopt the definition of power spectrum in Eq.(\ref{eq:pow}), but with a $1/10^5$ normalization. The results are averaged over 100 realizations. When investigating the effect on the power spectrum due to one defect at different positions along the lattice, we measure the power spectra for both the entire system $I_{TOT}(\omega)$ and the left and right subsystems separated by the defect, $I_{L}(\omega)$ and $I_{R}(\omega)$. As in previous studies, oscillations are found in the LD/HD phases. However unexpected and intriguing results emerge when the location ($  k$) and the ``strength'' ($q$) of the defect are varied. We shall elaborate these results in the following sections. 
  
 \subsection{Defect off center: $k \neq \mathbb{L}/2$}

\begin{figure}[h]
\begin{center}
\includegraphics[width=10cm,height=8cm]{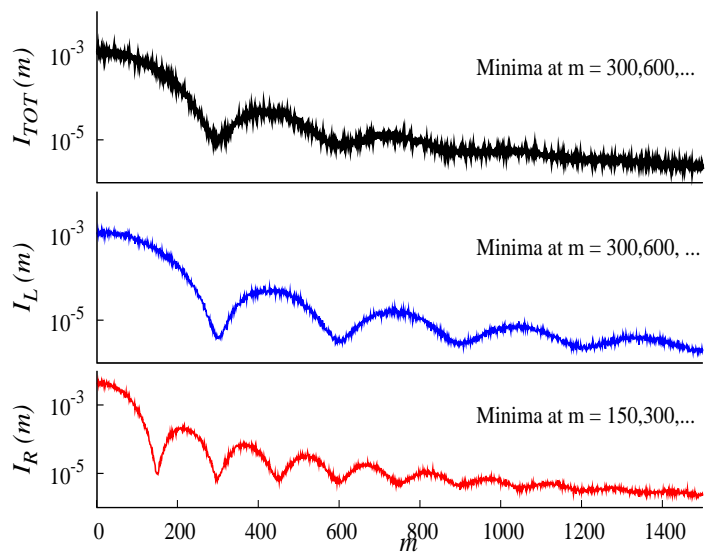}
\hspace{-2cm}
\caption{Power spectra of the entire lattice (top, black online), left (middle, blue) and right (bottom, red) sublattices with $\mathbb{L}=1000$, $k=333$ and $q=0.001$. $\omega = 2\pi m/T$.}
\label{fig:offcenter}
\end{center}
\end{figure}

\ \ \ \ Let us first introduce a single defect of hopping rate $q$ at site $k\neq \mathbb{L}/2$. The average bulk density of the system becomes:
\begin{equation}
\rho_{TOT}=\frac{k}{\mathbb{L}}\rho_{L}+\frac{\mathbb{L}-k}{\mathbb{L}}\rho_{R}.
\end{equation}

When the defect is off-center, $\rho_{TOT}$ is found in either HD or LD phase, while the left sublattice is in HD with $\rho_{L}= 1-q_{\textrm{\tiny{eff}}}$ and right sublattice in LD with $\rho_{R}=q_{\textrm{\tiny{eff}}}$. 
Once a hole-particle pair is created at site $k$, it introduces fluctuations in the hole density to the left and particle density to the right. As the defect is off center, the fluctuations in the left propagate with $v_{L}=1-2q_{\textrm{\tiny{eff}}}\equiv v$ for a distance of $k$ and the fluctuations in the right with $v_{R}=v$ for a distance of $(\mathbb{L}-k)$. We therefore expect the minima in $I_{L}(\omega)$  and $I_{R}(\omega)$  to be at $n\cdot vT/k$ and $n\cdot vT/(\mathbb{L}-k)$, respectively, since each sublattice may be treated as an independent TASEP as in \cite{AZS}. Concerning the entire lattice, on the other hand, the power spectrum measures the difference in timescale between the hole and particle fluctuations propagating through the left and right subsystems. The particle and hole fluctuations propagating through the sublattices do not cause a fluctuation in the overall particle density; only when one of the fluctuations remains following its counterpart leaving the system does the total lattice density change.  Therefore, the minima of $I_{TOT}(\omega)$ are located at $n\cdot vT/(\mathbb{L}_L-\mathbb{L}_R)=n\cdot vT/(\mathbb{L}-2k)$. 

Fig.\ref{fig:offcenter} displays the simulation result when $k=\mathbb{L}/3=333$ and $q=0.001$. It is reasonable to take $q_{\textrm{\tiny{eff}}}\approx q=0.001$, yielding $v =0.998$.  As expected, both $I_{L}(\omega)$ and $I_{R}(\omega)$ demonstrate oscillations of which the minima well characterize the subsystem sizes:
\begin{eqnarray*}
\textrm{Minima of left sublattice:} &=& n\cdot (0.998)10^5/333 \approx n\cdot 300\\
\textrm{Minima of right sublattice:} &=& n\cdot (0.998)10^5/666 \approx n\cdot 150
\end{eqnarray*}
We also observe oscillations in $I_{TOT}(\omega)$. In this case, $|\mathbb{L}_L-\mathbb{L}_R| = \mathbb{L}_L$ and hence we expect the minima in $I_{TOT}(\omega)$ to be at the same location as those of $I_{L}(\omega)$, which is precisely captured in Fig. \ref{fig:offcenter}.

\subsection{Intermediate range of $\omega$}
\begin{figure}[h]
\begin{center}
\includegraphics[width=14cm,height=10cm]{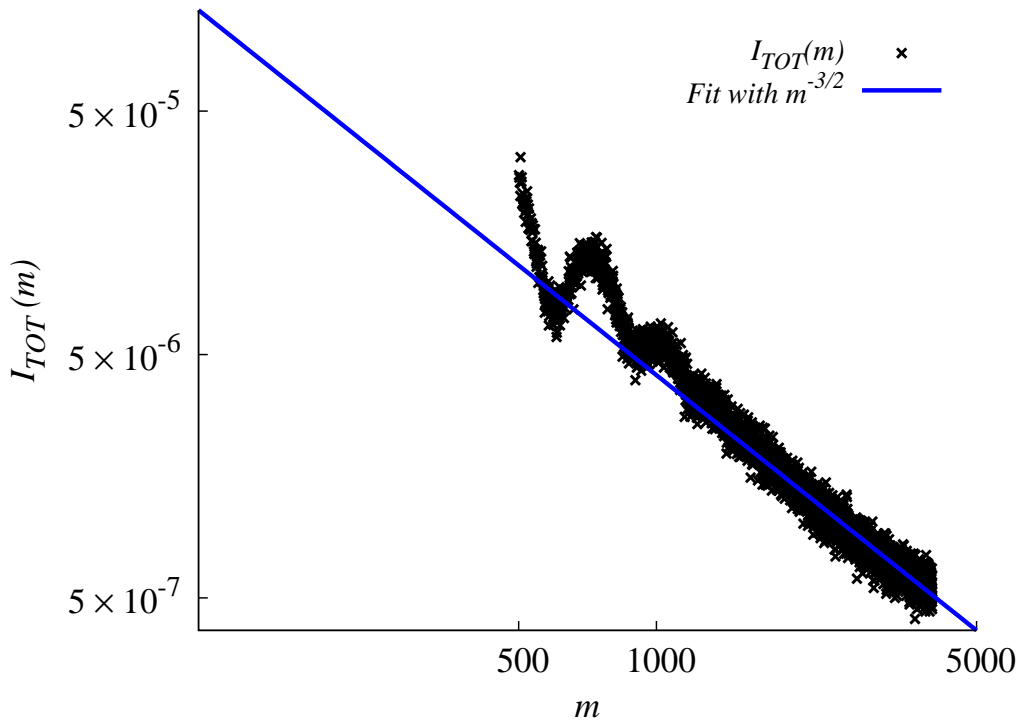}
\hspace{-2cm}
\caption{$I_{TOT}(m)$ with $m \in [500, 4000]$ for $\mathbb{L}=1000$, $k=333$ and $q=0.001$. $\omega = 2\pi m/T$. }
\label{fig:fit_o}
\end{center}
\end{figure}
\ \ \ \ In addition to the previous simulation findings, we want to comment on, without going into much detail, the behavior of $I_{TOT}(\omega)$ demonstrated in the intermediate range of $\omega$, i.e. $1 \ll m \ll T$ when oscillations are present. Consistent with the simulation and theory results in \cite{AZS}, here we retrieved the power law decay in the power spectrum of the entire system. For a system size of 1000, an exponent of -1.5 decently captures the decay (Fig.\ref{fig:fit_o}). Upon closer examination, the rate of decay varies with the system size. As discussed in \cite{AZS}, the exponent approaches -2 if the system is large enough to allow the oscillations to  damp completely. Our simulation confirms this conclusion. The continuum approach adopted in \cite{AZS}, on the other hand, is not suitable to analyze the system with a defect. Our theoretical treatment turns to the discrete space and time units for the low-frequency regime. We leave the intermediate $\omega$ regime for future pursuit.

 \subsection{Defect at center:$k = \mathbb{L}/2$}
 
\begin{figure}[h]
\begin{center}
\includegraphics[width=14cm,height=10cm]{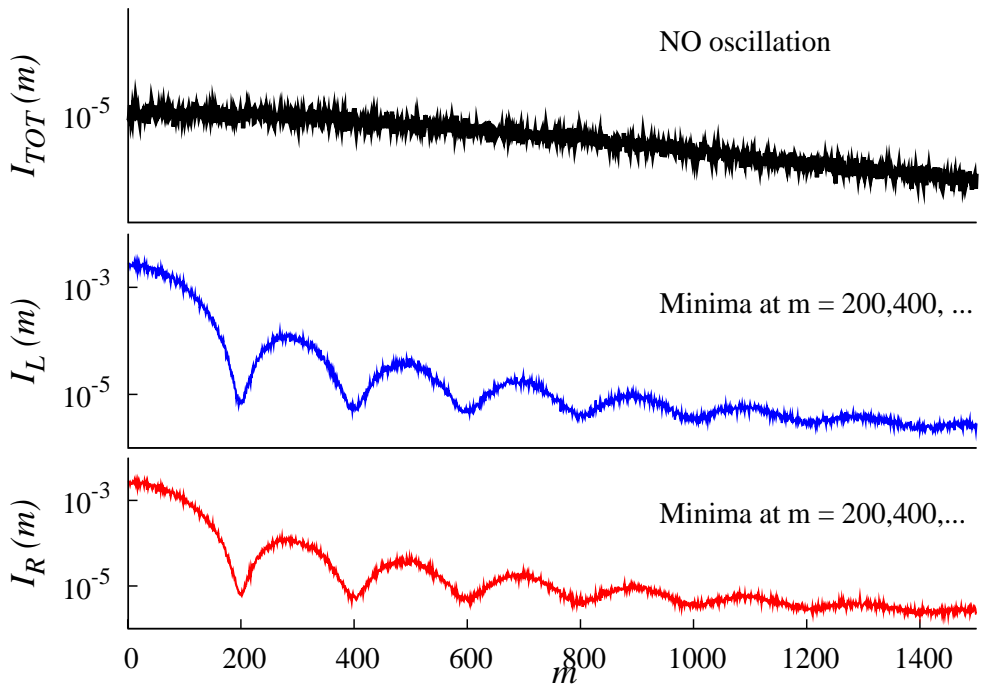}
\caption{Power spectra of the entire lattice (top, black online), left (middle, blue) and right (bottom, red) sublattices with $\mathbb{L}=1000$, $k=500$ and $q=0.001$. $\omega = 2\pi m/T$.}
\label{fig:center}
\end{center}
\end{figure}

\ \ \ \ Now we turn to a special case when the defect is located at the center of the system($k=\mathbb{L}/2$). Due to particle-hole symmetry, the system is always half-filled ($\rho_{TOT}=1/2$). The left and right sublattices again are in HD and LD, respectively and the minima in the power spectra are at the same location:$n\cdot vT/k$. When looking at $I_{TOT}(\omega)$, what behavior shall one expect when $\mathbb{L}_L-\mathbb{L}_R=0$? Now that the hole-particle pair is created right in the middle, it takes \emph{on average} the same time for the hole and particle to exit the system. However, it is the fluctuation of the timescale that $I_{TOT}(\omega)$ tries to capture.

Through our simulation data in Fig.\ref{fig:center} where $k=\mathbb{L}/2=500$ and $q=0.001$, we again retrieve the oscillations in $I_{L}(\omega)$ and $I_{R}(\omega)$ of which the minima are both at:
\begin{eqnarray*}
 n\cdot (0.998)10^5/500 \approx n\cdot 200
\end{eqnarray*}
However \emph{no oscillation} is observed in $I_{TOT}(\omega)$. The origin of such behavior is explored in Section  \ref{sec:theo}. 

\section{\label{sec:theo}Theory}

\ \ \ \ In this section, we present an explanation for the previously discussed simulation results.  Our approach utilizes a Langevin equation for the site density $\rho(x,t)$, which we linearize about its average $\bar{\rho}(x)$ \cite{AZS,CZ}.  In addition, the theory uses discrete space and time to naturally include UV cutoffs, which the authors of \cite{CZ} previously used for the constrained TASEP.  Since mRNA's typically contain several hundred codons, we avoid taking the continuum limit that was used previously in \cite{AZS}.  Further, we do not take the Fourier transform of the spatial component as in \cite{AZS} and \cite{CZ} due to the breaking of translational invariance by the slow site.  Instead, we solve a set of $\mathbb{L}$ linear equations obtained from the Langevin equation for each site.  We also choose the timescale to be a single update attempt with $T$ being the number of updates in a run as in \cite{CZ}.  This choice is in contrast with what is chosen for the simulation, where data is taken every 10 MCS.  To account for the difference, we must sum over the harmonics \cite{CZ} of our theoretical results to make a meaningful comparison.

Using a similar approach as \cite{CZ}, we begin with the Langevin equation for the site density:
\begin{equation}
\fl\partial_t \rho(x,t)=\frac{\rho(x-1,t)[1-\rho(x,t)]-\rho(x,t)[1-\rho(x+1,t)]}{\mathbb{L}+1}+\xi(x-1,t)-\xi(x,t)
\end{equation}
where $x\notin\{1,k,k+1,\mathbb{L}\}$ and time is on the order of an update.  For those sites, we have:
\begin{eqnarray}
\partial_t \rho(1,t)&=&\frac{\alpha[1-\rho(1,t)]-\rho(1,t)[1-\rho(2,t)]}{\mathbb{L}+1}+\xi(0,t)-\xi(1,t)\\
\partial_t \rho(k,t)&=&\frac{\rho(k-1,t)[1-\rho(k,t)]-q\rho(k,t)[1-\rho(k+1,t)]}{\mathbb{L}+1}\\
\nonumber &&+\xi_(k-1,t)-\xi(k,t)\\
\partial_t \rho(k+1,t)&=&\frac{q\rho(k,t)[1-\rho(k+1,t)]-\rho(k+1,t)[1-\rho(k+2,t)]}{\mathbb{L}+1}\\
\nonumber&&+\xi(k,t)-\xi(k+1,t)\\
\partial_t \rho(\mathbb{L},t)&=&\frac{\rho(\mathbb{L}-1,t)[1-\rho(\mathbb{L},t)]-\beta\rho(\mathbb{L},t)[1-\rho(\mathbb{L},t)]}{\mathbb{L}+1}\\
\nonumber &&+\xi(\mathbb{L}-1,t)-\xi(\mathbb{L},t)
\end{eqnarray}
where $q$ is the hopping rate for the slow site and $\xi$ is assumed to be homogeneous, uncorrelated white noise associated with hopping from site $x$ to $x+1$ with zero mean and variance $A$ such that $\langle\xi(x,t)\xi(x^\prime,t^\prime)\rangle=A\delta_{x,x^\prime}\delta_{t,t^\prime}$ \cite{CZ}. This assumption for the noise will later be validated by the agreement of our analytical result with the simulation results.  The $1/(\mathbb{L}+1)$ on the right-hand side is due to the probability of choosing a particular site to update.

To simplify the calculation, we choose $\alpha$ and $\beta$ such that the profile is flat (i.\ e.\ no $x$ dependence) for the two regions separated by the slow site.  This is also consistent with the choice of parameters in the simulations.  Thus, we have
\begin{eqnarray}
\bar{\rho}(x)=\bar{\rho}_L \quad & x\in[1,k];&
\bar{\rho}_L=\frac{1}{1+q}=\alpha;\\
\bar{\rho}(x)=\bar{\rho}_R  &x\in[k+1,L]; \quad&
\bar{\rho}_R=\frac{q}{1+q}=\beta
\end{eqnarray}
Note, $\bar{\rho}_L+\bar{\rho}_R=1$ with this choice of parameters.  Particle/hole symmetry emerges with this choice of parameters as $\bar{\rho}_L+\bar{\rho}_R=1$.  We then define the fluctuations of $\rho(x,t)$ about its average to be
\begin{eqnarray}
\varphi(x,t)=&\rho(x,t)-\bar{\rho}_L \quad x\in[1,k]\\
\varphi(x,t)=&\rho(x,t)-\bar{\rho}_R \quad x\in[k+1,L]
\end{eqnarray}

For small values of $q$, the particle spacing on the right sublattice is so large that the interactions do not contribute significantly.  Similarly on the left sublattice by particle/hole symmetry, the hole spacing should be large enough that they do not interact with each other.  These interactions appear as quadratic terms in $\varphi(x,t)$, which we will assume to be small and dropped in the following.

At this point, we will treat the left and right sublattices separately.  Rewriting the Langevin equations and keeping only terms linear in $\varphi$, we have
\begin{eqnarray}\label{Langevin-phi}
\fl\partial_t\varphi(1,t)=\frac{-\varphi(1,t)+\bar{\rho}_L\varphi(2,t)}{\mathbb{L}+1}+\xi(0,t)-\xi(1,t)\\
\fl\partial_t \varphi(x,t)=\frac{(1-\bar{\rho}_L)\varphi_L(x-1,t)-\varphi_L(x,t)+\bar{\rho}_L\varphi_L(x+1,t)}{\mathbb{L}+1}+\xi(x-1,t)-\xi(x,t)\\
x\in[2,k-1]\nonumber\\
\fl\partial_t \varphi(k,t)=\frac{(1-\bar{\rho}_L)\varphi(k-1,t)-\varphi(k,t)+\bar{\rho}_R\varphi(k+1,t)}{\mathbb{L}+1}+\xi(k-1,t)-\xi(k,t)\\
\fl\partial_t \varphi(k+1,t)=\frac{(1-\bar{\rho}_L)\varphi(k,t)-\varphi(k+1,t)+\bar{\rho}_R\varphi(k+2,t)}{\mathbb{L}+1}+\xi(k,t)-\xi(k+1,t)\\
\fl\partial_t \varphi(x,t)=\frac{(1-\bar{\rho}_R)\varphi(x-1,t)-\varphi(x,t)+\bar{\rho}_R\varphi(x+1,t)}{\mathbb{L}+1}+\xi(x-1,t)-\xi(x,t)\\
x\in[k+2,\mathbb{L}-1]\nonumber\\
\fl\partial_t \varphi(\mathbb{L},t)=\frac{(1-\bar{\rho}_R)\varphi(\mathbb{L}-1,t)-\varphi(\mathbb{L},t)}{\mathbb{L}+1}+\xi(\mathbb{L}-1,t)-\xi(\mathbb{L},t)
\end{eqnarray}
Taking the Fourier transform to be
\begin{eqnarray}
\varphi(x,t)=\sum_\omega e^{-i\omega t}\tilde{\varphi}(x,\omega)\\
\tilde{\varphi}(x,\omega)=\frac{1}{T}\sum_t e^{i\omega t}\varphi(x,t)
\end{eqnarray}
with
\begin{equation}
\omega=\frac{2\pi m}{T} \quad m\in[1,T]
\end{equation}
we arrive at
\begin{eqnarray}\label{FT-phi}
\zeta(\omega)\tilde{\varphi}(1,\omega)-\frac{\bar{\rho}_L}{\mathbb{L}+1}\tilde{\varphi}(2,\omega)=\tilde{\eta}(1,\omega)\\
-\frac{1-\bar{\rho}_L}{\mathbb{L}+1}\tilde{\varphi}(x-1,\omega)+\zeta(\omega)\tilde{\varphi}(x,\omega)-\frac{\bar{\rho}_L}{\mathbb{L}+1}\tilde{\varphi}(x+1,\omega)=\tilde{\eta}(x,\omega)\\
\hspace{21em} x\in[2,k-1]\nonumber\\
-\frac{1-\bar{\rho}_L}{\mathbb{L}+1}\tilde{\varphi}(k-1,\omega)+\zeta(\omega)\tilde{\varphi}(k,\omega)-\frac{\bar{\rho}_R}{\mathbb{L}+1}\tilde{\varphi}(k+1,\omega)=\tilde{\eta}(k,\omega)\\
-\frac{1-\bar{\rho}_L}{\mathbb{L}+1}\tilde{\varphi}(k,\omega)+\zeta(\omega)\tilde{\varphi}(k+1,\omega)-\frac{\bar{\rho}_R}{\mathbb{L}+1}\tilde{\varphi}(k+2,\omega)=\tilde{\eta}(k+1,\omega)\\
-\frac{1-\bar{\rho}_R}{\mathbb{L}+1}\tilde{\varphi}(x-1,\omega)+\zeta(\omega)\tilde{\varphi}(x,\omega)-\frac{\bar{\rho}_R}{\mathbb{L}+1}\tilde{\varphi}(x+1,\omega)=\tilde{\eta}(x,\omega)\\
\hspace{21em} x\in[k+2,\mathbb{L}-1]\nonumber\\
-\frac{1-\bar{\rho}_R}{\mathbb{L}+1}\tilde{\varphi}(\mathbb{L}-1,\omega)+\zeta(\omega)\tilde{\varphi}(\mathbb{L},\omega)=\tilde{\eta}(\mathbb{L},\omega)
\end{eqnarray}
where
\begin{eqnarray}
\zeta(\omega)=e^{i\omega}-1+\frac{1}{\mathbb{L}+1}\\
\tilde{\eta}(x,\omega)=\tilde{\xi}(x-1,\omega)-\tilde{\xi}(x,\omega)
\end{eqnarray}
or written more compactly,
\begin{equation}
\mathbb{M}\tilde{\varphi}=\tilde{\eta}
\end{equation}
where $\tilde{\varphi}$ and $\tilde{\eta}$ are vectors in $x$.  The $\mathbb{L}\times\mathbb{L}$ tridiagonal matrix $\mathbb{M}$ has entries:
\begin{eqnarray}
M_{ii}=\zeta(\omega)\\
M_{ii+1}=-\frac{\bar{\rho}_L}{\mathbb{L}+1} \quad i \in[1,k-1]\\
M_{ii+1}=-\frac{\bar{\rho}_R}{\mathbb{L}+1}  \quad i \in[k,\mathbb{L}-1]\\
M_{ii-1}=-\frac{\bar{\rho}_R}{\mathbb{L}+1} \quad i \in[2,k+1]\\
M_{ii-1}=-\frac{\bar{\rho}_L}{\mathbb{L}+1}  \quad i \in[k+2,\mathbb{L}]
\end{eqnarray}
where we have made use of $\bar{\rho}_L+\bar{\rho}_R=1$.  By inverting the matrix ($\mathbb{M}^{-1}=\mathbb{S}$), we have the $\tilde{\varphi}$'s in terms of the $\tilde{\eta}$'s, namely
\begin{equation}
\tilde{\varphi}(x,\omega)=\sum_{y=1}^\mathbb{L} \mathbb{S}_{xy}\tilde{\eta}(y,\omega)
\end{equation}
The inverse of this type of tridiagonal matrix is given in \cite{Huang97}.

We are interested in power spectrum of $I=\langle|\tilde{N}|^2\rangle$, so first we need $\tilde{N}$.  $\tilde{N}$ is related to $\tilde{\varphi}$ by summing over the spatial coordinate.  Thus from \eref{FT-phi}, we have
\begin{eqnarray}
\fl\tilde{N}_{TOT}(\omega)=\frac{\frac{\bar{\rho}_L}{\mathbb{L}+1}\left[\tilde{\varphi}(1,\omega)+\tilde{\varphi}(\mathbb{L},\omega)\right]+\tilde{\xi}(\mathbb{L},\omega)-\tilde{\xi}(0,\omega)}{1-e^{i\omega}}\\
\fl\tilde{N}_L(\omega)=\frac{\frac{1}{\mathbb{L}+1}\left[\bar{\rho}_R\tilde{\varphi}(k+1,\omega)-\bar{\rho}_R\tilde{\varphi}(k,\omega)-\bar{\rho}_L\tilde{\varphi}(1,\omega)\right]+\tilde{\xi}(0,\omega)-\tilde{\xi}(k,\omega)}{e^{i\omega}-1}\\
\fl\tilde{N}_R(\omega)=\frac{\frac{1}{\mathbb{L}+1}\left[\bar{\rho}_R\tilde{\varphi}(k,\omega)-\bar{\rho}_R\tilde{\varphi}(k+1,\omega)-\bar{\rho}_L\tilde{\varphi}(\mathbb{L},\omega)\right]+\tilde{\xi}(k,\omega)-\tilde{\xi}(\mathbb{L},\omega)}{e^{i\omega}-1}
\end{eqnarray}
for the entire lattice and the left and right sublattices, respectively.  Upon squaring $\tilde{N}$ and averaging over the noise to calculate the power spectra, we still need to sum over the harmonics to make meaningful comparisons with the simulation data \cite{CZ}.  The harmonic sum is necessary since the simulation data is not recorded at every update.  If we take data every $\ell$ updates, then
\begin{equation}
I(m)=\sum_{z=0}^{\ell-1}I(\omega_{m,z})
\end{equation}
where
\begin{equation}
\omega_{m,z}=\frac{2\pi}{T}\left(m+\frac{zT}{\ell}\right)
\end{equation}
Finally, the only remaining parameter in the theory is the variance of the noise $A$.  Instead of leaving it as a fit parameter, $A$ is obtained from the microscopic description \cite{PPOF}
\begin{equation}
A=\frac{q}{(\mathbb{L}+1)(1+q)^2}
\end{equation}
which is simply the current divided by $\mathbb{L}+1$.  Thus, the final result is
\begin{eqnarray}
\fl I_{TOT}(m)=\sum_{z=0}^{\ell-1}\frac{q}{(\mathbb{L}+1)(1+q)^2T(2-2\cos\omega_{m,z})}\label{I-tot}\\
\times\left[\left(\frac{\bar{\rho}_L}{\mathbb{L}+1}\right)^2\left\{\langle\tilde{\varphi}(1,\omega_{m,z})\tilde{\varphi}^*(1,\omega_{m,z})\rangle+\langle\tilde{\varphi}(\mathbb{L},\omega_{m,z})\tilde{\varphi}^*(\mathbb{L},\omega_{m,z})\rangle\right.\right.\nonumber\\
+\left. 2\Re\left[\langle\tilde{\varphi}(1,\omega_{m,z})\tilde{\varphi}^*(\mathbb{L},\omega_{m,z})\rangle\right]\right\}\nonumber\\
+\left.\left(\frac{2\bar{\rho}_L}{\mathbb{L}+1}\right)\Re\left\{\langle\left[\tilde{\varphi}(1,\omega_{m,z})+\tilde{\varphi}(\mathbb{L},\omega_{m,z})\right]\left[\tilde{\xi}^*(\mathbb{L},\omega_{m,z})-\tilde{\xi}^*(0,\omega_{m,z})\right]\rangle\right\}+2\right]\nonumber\\
\fl I_{L}(m)=\sum_{z=0}^{\ell-1}\frac{q}{(\mathbb{L}+1)(1+q)^2T(2-2\cos\omega_{m,z})}\label{I-L}\\
\times\left[\left(\frac{\bar{\rho}_R}{\mathbb{L}+1}\right)^2\left\{\langle\tilde{\varphi}(k,\omega_{m,z})\tilde{\varphi}^*(k,\omega_{m,z})\rangle+\langle\tilde{\varphi}(k+1,\omega_{m,z})\tilde{\varphi}^*(k+1,\omega_{m,z})\rangle\right.\right.\nonumber\\
-\left.2\Re\left[\langle\tilde{\varphi}(k,\omega_{m,z})\tilde{\varphi}^*(k+1,\omega_{m,z})\rangle\right]\right\}+\left(\frac{\bar{\rho}_L}{\mathbb{L}+1}\right)^2\langle\tilde{\varphi}(1,\omega_{m,z})\tilde{\varphi}^*(1,\omega_{m,z})\rangle\nonumber\\
+\left[\frac{2\bar{\rho}_L\bar{\rho}_R}{(\mathbb{L}+1)^2}\right]\Re\left\{\langle\left[\tilde{\varphi}(k,\omega_{m,z})-\tilde{\varphi}(k+1,\omega_{m,z})\right]\tilde{\varphi}^*(1,\omega_{m,z})\rangle\right\} \nonumber\\
+\left(\frac{2\bar{\rho}_R}{\mathbb{L}+1}\right)\Re\left\{\langle\left[\tilde{\varphi}(k,\omega_{m,z})-\tilde{\varphi}(k+1,\omega_{m,z})\right]\left[\tilde{\xi}^*(k,\omega_{m,z})-\tilde{\xi}^*(0,\omega_{m,z})\right]\rangle\right\}\nonumber\\
+\left.\left(\frac{2\bar{\rho}_L}{\mathbb{L}+1}\right)\Re\left\{\langle\tilde{\varphi}(1,\omega_{m,z})\left[\tilde{\xi}^*(k,\omega_{m,z})-\tilde{\xi}^*(0,\omega_{m,z})\right]\rangle\right\}+2\right]\nonumber\\
\fl I_{R}(m)=\sum_{z=0}^{\ell-1}\frac{q}{(\mathbb{L}+1)(1+q)^2T(2-2\cos\omega_{m,z})}\label{I-R}\\
\times\left[\left(\frac{\bar{\rho}_R}{\mathbb{L}+1}\right)^2\left\{\langle\tilde{\varphi}(k,\omega_{m,z})\tilde{\varphi}^*(k,\omega_{m,z})\rangle+\langle\tilde{\varphi}(k+1,\omega_{m,z})\tilde{\varphi}^*(k+1,\omega_{m,z})\rangle\right.\right.\nonumber\\
-\left.2\Re\left[\langle\tilde{\varphi}(k,\omega_{m,z})\tilde{\varphi}^*(k+1,\omega_{m,z})\rangle\right]\right\}+\left(\frac{\bar{\rho}_L}{\mathbb{L}+1}\right)^2\langle\tilde{\varphi}(\mathbb{L},\omega_{m,z})\tilde{\varphi}^*(\mathbb{L},\omega_{m,z})\rangle\nonumber\\
+\left[\frac{2\bar{\rho}_L\bar{\rho}_R}{(\mathbb{L}+1)^2}\right]\Re\left\{\langle\left[\tilde{\varphi}(k+1,\omega_{m,z})-\tilde{\varphi}(k,\omega_{m,z})\right]\tilde{\varphi}^*(\mathbb{L},\omega_{m,z})\rangle\right\} \nonumber\\
+\left(\frac{2\bar{\rho}_R}{\mathbb{L}+1}\right)\Re\left\{\langle\left[\tilde{\varphi}(k,\omega_{m,z})-\tilde{\varphi}(k+1,\omega_{m,z})\right]\left[\tilde{\xi}^*(k,\omega_{m,z})-\tilde{\xi}^*(\mathbb{L},\omega_{m,z})\right]\rangle\right\}\nonumber\\
+\left.\left(\frac{2\bar{\rho}_L}{\mathbb{L}+1}\right)\Re\left\{\langle\tilde{\varphi}(\mathbb{L},\omega_{m,z})\left[\tilde{\xi}^*(\mathbb{L},\omega_{m,z})-\tilde{\xi}^*(k,\omega_{m,z})\right]\rangle\right\}+2\right]\nonumber
\end{eqnarray}
where
\begin{eqnarray}
\langle\tilde{\varphi}(a,\omega)\tilde{\varphi}^*(b,\omega)\rangle=2\sum_{x=1}^\mathbb{L}\mathbb{S}_{ax}\mathbb{S}^*_{bx}-\sum_{x=2}^\mathbb{L}\left[\mathbb{S}_{ax}\mathbb{S}^*_{b(x-1)}+\mathbb{S}_{a(x-1)}\mathbb{S}^*_{bx}\right]\\
\langle\tilde{\varphi}(a,\omega)\tilde{\xi}^*(0,\omega)\rangle=\mathbb{S}_{a1}\\
\langle\tilde{\varphi}(a,\omega)\tilde{\xi}^*(\mathbb{L},\omega)\rangle=-\mathbb{S}_{a\mathbb{L}}\\
\langle\tilde{\varphi}(a,\omega)\tilde{\xi}^*(b,\omega)\rangle=\mathbb{S}_{a(b+1)}-\mathbb{S}_{ab}; \quad b\neq 0,\mathbb{L}
\end{eqnarray}
and the dependence of $\mathbb{S}$ on $\omega$ has been suppressed.

Before making a comparison with the simulation results, we will focus on the special case of $k=\mathbb{L}/2$.  In this case, both sublattices have the same lengths, while we still have the particle/hole symmetry previously mentioned.  Using these symmetries, we reduce the number of unique elements in the $\mathbb{S}$ matrix.  Specifically, we have
\begin{equation}
\mathbb{S}_{ij}=\mathbb{S}_{(\mathbb{L}-i+1)(\mathbb{L}-j+1)}
\end{equation}

Further, the cross correlation between the first and last site $\langle\tilde{\varphi}(1,\omega)\tilde{\varphi}_\mathbb{L}^*\rangle$, which only appears in the power spectrum of the entire lattice, is a real quantity. The cross correlations in the sublattices' power spectra are not real.  In the previously studied homogeneous TASEP, the oscillations appeared mathematically as the real part of a complex function \cite{AZS}.  Here, it is less clear how the oscillations emerge, as opposed to the simpler expression in \cite{AZS}.  However, we speculate that the real correlation terms are the source of the disappearance of the oscillation, and we are currently investigating this idea.

\begin{figure}[htb]
\begin{center}
\includegraphics[width=0.7\textwidth]{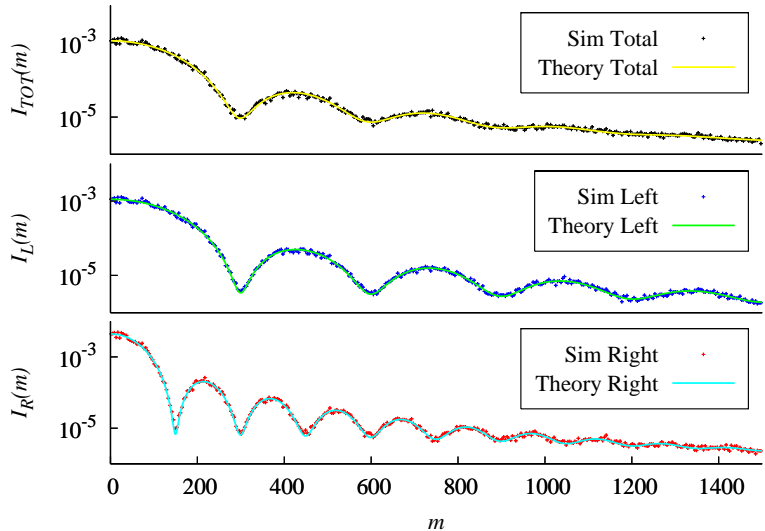}
\end{center}
\caption{Comparison of simulation and theoretical power spectra for $\mathbb{L}=1000$, $k=333$, and $q=0.001$.\label{k333-q001}}
\end{figure}

\begin{figure}[htb]
\begin{center}
\includegraphics[width=0.7\textwidth]{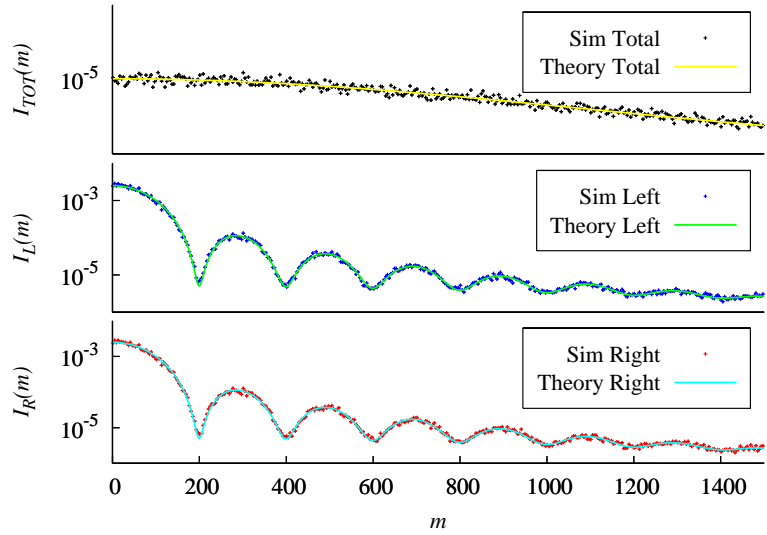}
\end{center}
\caption{Comparison of simulation and theoretical power spectra for $\mathbb{L}=1000$, $k=500$, and $q=0.001$.\label{k500-q001}}
\end{figure}

Using equations \eref{I-tot}, \eref{I-L}, and  \eref{I-R}, we compare the theoretical result to the simulation data.  For low $q$ values, the agreement is remarkably good.  In figure \ref{k333-q001}, we show the comparison for $\mathbb{L}=1000$, $k=333$, and $q=0.001$.  
As shown in the figure, the oscillations for all three power spectra are correctly predicted for the entire range of $m$.  The theory also predicts the vanishing of the oscillations for the total power spectrum, as shown in figure \ref{k500-q001}, for $\mathbb{L}=1000$, $k=500$, and $q=0.001$.  
Again, the theory correctly predicts the oscillations of the two sublattices with very good agreement.  Therefore, the theory incorporates all the necessary information needed to accurately reproduce the power spectra for small $q$ values.

\begin{figure}[htb]
\begin{center}
\includegraphics[width=0.7\textwidth]{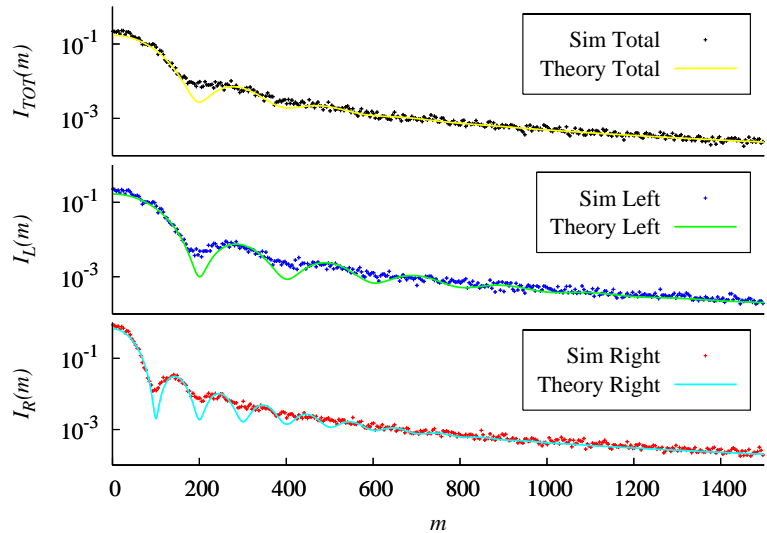}
\end{center}
\caption{Comparison of simulation and theoretical power spectra for $\mathbb{L}=1000$, $k=333$, and $q=0.2$.\label{k333-q20}}
\end{figure}

As we increase the $q$ value, a difference between the simulation data and the theoretical result begins to appear.  This difference is shown in figure \ref{k333-q20} for $\mathbb{L}=1000$, $k=333$, and $q=0.2$.  
While the theory does correctly predict the $m$ value of the minima, the damping of the oscillations is greater in the simulation than in the theory.  Physically, the particles/holes begin to interact more frequently (via the exclusion) at larger $q$ values.  However, we employ a linear theory void of these interactions.  The effect of the interactions is an enhancement of the effect diffusion coefficient \cite{AZS,AZ09}, which controls the damping of the oscillations \cite{AZS,AZ09}. Therefore, it is no surprise that the theory is not in perfect agreement with the simulation data.  Since the interactions terms (terms quadratic in $\varphi$) are assumed to be small, perturbation theory should be utilized to account for the difference, not only here but for the homogeneous TASEP \cite{AZS} and other modified TASEP models \cite{CZ} as well.

\section{\label{sec:sum}Summary and outlook}
\ \ \ \ In this paper, we explored the power spectrum associated with the total occupancy of a TASEP with a single slow site.  The slow site (at position $k$ with rate $q$) served as a bottleneck for particles passing through from the left to the right sublattice.  Using Monte Carlo simulations, we observed oscillations, similar to those seen in the HD/LD phases of the homogeneous TASEP \cite{AZS}, for the two sublattices separated by the slow site and for the entire lattice.  However, the oscillations for the total power spectrum vanished when the slow site was place in the center of the lattice, while the oscillations remained for the sublattices.  A theoretical approach utilizing a linear Langevin equation (with discrete space and time) for the density correctly captured this disappearance for $k=\mathbb{L}/2$ as well as the oscillations for the other cases.  The theory had excellent agreement when compared to the simulation results when the hopping rate for the slow site $q$ was small.  As the rate was increased, the overall agreement was not as impressive, but the locations of the minima of the oscillations for the simulation and theory were the same.

While the theoretical result presented is not very transparent, we can glean some insight into the disappearance of the oscillations for the power spectrum of the entire system when $k=\mathbb{L}/2$.  Specifically, we found the cross correlation between the density fluctuations of the first and last site is a real quantity, as opposed to having an imaginary component as well.  For the homogeneous TASEP \cite{AZS}, the oscillations emerged from the real part of a complex function which we lack in this special case for the single slow site in the center.  We speculate that the cross correlation being a real function signals the disappearance of the oscillations.

Open questions remain beyond this study.  One immediate question is the connection between our theoretical approach and the one in \cite{AZS}.  While there is good agreement between our theory and simulation results, the theory lacks the simple, intuitive explanation demonstrated in \cite{AZS}.  By taking the appropriate limits, we would like to have a similar simple expression for the oscillations that can be used to show the $\mathbb{L}$ and $k$ dependence of the minima for the power spectrum of the entire system, especially in the $k=\mathbb{L}/2$ case.  Finding such a connection would not only expand our understand for a slow site, but other extensions \cite{CZS09,DSZ07} as well.  Another question pertains to the power law decay found in the $k=\mathbb{L}/2$ case power spectrum for the entire system.  The simulation data shows a $\omega^{-3/2}$ decay after some crossover scale.  We wish to explore the $\mathbb{L}$ dependence of the crossover as well as explore the physics behind the power law.  Lastly, what is the effect of two or more slow sites \cite{DSZ06,TomChou} and extended objects \cite{DSZ07,Chou03} on the power spectrum?

Many biological extensions of this model exist that would more closely realize what happens in nature.  For the translation process, these extensions include coupling a single slow site with a finite pool of particles \cite{ASZ08} to more accurately model the finite resources in the cell and multiple TASEPs \cite{CZS09} to capture the effects of competition for resources in a cell.  Also, one should have completely inhomogeneous hopping rates to model the various codon elongation rates.  By studying the individual elements and their coupled effects, we gain a better understanding of the translation process in protein synthesis.  Outside of the context of protein synthesis, many other biological systems may be studied through the perspective of physics.  From social networks to predator-prey dynamics, much work is left to be done to understand nature around us.

\section*{Acknowledgement}
The authors would like to thank Royce K.P. Zia and Beate Schmittmann for stimulating discussions and critical suggestion on theoretical analyses, as well as their hospitality hosting J.J.D. at Virginia Tech where some of the work was performed. This research is supported in part by U.S. National Science Foundation through Grant No. DMR- 0705152.   This work is also in part supported by Washington and Lee University and Hamline University.

\section*{References}
\bibliographystyle{iopart-num} 
\bibliography{references}

\end{document}